\documentclass[10pt]{iopart}

\usepackage{graphicx}
\usepackage{hyperref}
\usepackage{lineno, color}
\usepackage{xcolor}
\begin{document}
\hypersetup{
pdftitle={In-situ coating of silicon-rich films on tokamak plasma-facing components with real-time Si material injection},
pdfsubject={plasma-material interactions},
pdfauthor={Florian Effenberg},
pdfkeywords={plasma facing components, material injection, siliconization, fusion materials, silicon oxide}
}

\title{In-situ coating of silicon-rich films on tokamak plasma-facing components with real-time Si material injection}

\author{F. Effenberg$^{1}$, S. Abe$^{1,2}$, G. Sinclair$^{3}$, T. Abrams$^{3}$, A. Bortolon$^{1}$, W.R. Wampler$^{4}$, F.M. Laggner$^{5}$, D.L. Rudakov$^{6}$, I. Bykov$^{3}$, C.J. Lasnier$^{7}$, D. Mauzey$^{1}$, A. Nagy$^{1}$, R. Nazikian$^{3}$, F. Scotti$^{7}$, H.Q. Wang$^{3}$, R.S. Wilcox$^{8}$, and the DIII-D Team}

\address{1 - Princeton Plasma Physics Laboratory, Princeton, NJ 08543, USA}
\address{2 - Princeton University, Princeton, NJ 08544, USA}
\address{3 - General Atomics, San Diego, CA 92186, USA}
\address{4 - Sandia National Laboratories, Albuquerque, NM 87111, USA}
\address{5 - North Carolina State University, Raleigh, NC 27695, USA}
\address{6 - University of California - San Diego, La Jolla, CA 92093, USA}
\address{7 - Lawrence Livermore National Laboratory, Livermore, CA 94550, USA}
\address{8 - Oak Ridge National Laboratory, Oak Ridge, TN 37831, USA}
 
\ead{feffenbe@pppl.gov}
\vspace{10pt}
\begin{indented}
\item[]March 2023
\end{indented}

\begin{abstract}
{Experiments have been conducted in the DIII-D tokamak to explore the in-situ growth of silicon-rich layers as a potential technique for real-time replenishment of surface coatings on plasma-facing components (PFCs) during steady-state long-pulse reactor operation. Silicon (Si) pellets of 1 mm diameter were injected into low- and high-confinement (L-mode and H-mode) plasma discharges with densities ranging from $3.9-7.5\times10^{19}$ m$^{-3}$ and input powers ranging from $5.5-9$ MW. The small Si pellets were delivered with the impurity granule injector (IGI) at frequencies ranging from 4-16 Hz corresponding to mass flow rates of $5-19$ mg/s ($1-4.2\times10^{20}$ Si/s) at cumulative amounts of up to 34 mg of Si per five-second discharge. Graphite samples were exposed to the scrape-off layer and private flux region plasmas through the divertor material evaluation system (DiMES) to evaluate the Si deposition on the divertor targets. The Si II emission at the sample correlates with silicon injection and suggests net surface Si-deposition in measurable amounts. Post-mortem analysis showed Si-rich coatings containing silicon oxides, of which SiO$_2$ is the dominant component. No evidence of SiC was found, which is attributed to low divertor surface temperatures. The in-situ and ex-situ analysis found that Si-rich coatings of at least $0.4-1.2$ nm thickness have been deposited at $0.4-0.7$ nm/s. The technique is estimated to coat a surface area of at least 0.94 m$^2$ on the outer divertor. These results demonstrate the potential of using real-time material injection to form Si-enriched layers on divertor PFCs during reactor operation.}
\end{abstract}
%
% Uncomment for keywords
\vspace{2pc}
\noindent{\it Keywords}: siliconization, plasma-facing components, erosion, silicon oxide, divertor, material migration, real-time coating
%
% Uncomment for Submitted to journal title message
%\submitto{\NF}
%
% Uncomment if a separate title page is required
%\maketitle
% 
% For two-column output uncomment the next line and choose [10pt] rather than [12pt] in the \documentclass declaration
\ioptwocol
%-------------------------------------------------------------------------------
%-------------------------------------------------------------------------------
\section{Introduction\label{sec:Intro}}
%-------------------------------------------------------------------------------
The realization of a pilot Fusion Power Plant (FPP) requires a viable material solution \cite{menard_2016, buttery_2021}. In \cite{stangeby_2022}, estimated net erosion rates for solid plasma-facing components (PFCs) have been reviewed across a spectrum of materials, from low atomic number (Z) elements such as boron and beryllium to high-Z elements like iron and tungsten. For the International Thermonuclear Experimental Reactor (ITER), these rates fall in the range of approximately 10 to a few 100 kilograms per annum. For subsequent fusion devices, including US compact fusion pilot plants or the China Fusion Test Reactor (CFTR), rates could reach between $10$ and $10,000$ kilograms per year. Only limited concentrations of high-Z materials (e.g., tungsten) can be tolerated in a burning thermonuclear core plasma due to radiative losses. Therefore, the use of relatively thin, consumable low-Z refractory materials such as ceramics are being re-considered as a potential solution for the main walls \cite{tanabe_2019, stangeby_2022}. Given these observations, it has been suggested that in high-duty-cycle deuterium-tritium (DT) tokamaks, the application of non-metallic low-Z refractory materials like ceramics - including graphite, silicon carbide, etc. - could serve as in situ replenishable claddings \cite{stangeby_2022}. These relatively thin layers could be used to coat substrates that possess neutron resistance, offering a potential solution for safeguarding the main walls of future reactors while minimizing the potential degradation of the confined plasma. However, assessing the viability of this proposal necessitates further information, much of which still needs to be made available as of now. Real-time in-situ replenishment techniques need to be developed to prolong the lifespan of plasma-facing components and to minimize costly reactor downtime for maintenance. Implementing such approaches will depend on a comprehensive understanding and data collection regarding material properties and plasma-material interactions under these intense conditions. 

Experiments at NSTX, ASDEX-U, DIII-D, EAST, WEST, KSTAR, LHD, and W7-X demonstrated the feasibility of controlling the wall conditions and plasmas-material interactions with real-time material injections \cite{mansfield_2009, bortolon_real-time_2019,  sun_real_2019, andruczyk_2020, bortolon_2020, gilson_wall_2021, shoji_2020, nespoli_2020, lunsford_characterization_2021, effenberg_2022, bodner_2022}. The in-situ growth of low-to-medium-Z coatings is key to reducing recycling and impurity sources in these cases. 

The formation of Si-rich layers induced by siliconization in modern tokamaks under high-performance, high heat flux plasma conditions have been shown to be an effective wall conditioning technique that can compete with boronization for lowering oxygen impurity levels, increasing plasma density limits, and energy confinement \cite{winter_1993, samm_plasma_1995, gong_2001, duan_2007}. Furthermore, silicon carbide (SiC) is a promising low-to-medium-Z material option for future plasma-facing components due to its low hydrogenic diffusivity, high-temperature strength, and mechanical resilience to neutron damage \cite{snead_2011, koyanagi_2018, rudakov_net_2020, sinclair_quantifying_2021, abrams_2021sic, sinclair_2022}. 

The experiments presented in this work have been conducted at the DIII-D tokamak, which has a strong research program on core-edge integration and advanced materials \cite{rudakov_2017, fenstermacher_2022} including plans to utilize siliconization and install new SiC wall tile arrays \cite{buttery_2018, zamperini_2022}. 

Experimental results from in-situ real-time growth of silicon-rich layers on plasma-facing components by small silicon pellet injection at the outboard midplane in DIII-D plasmas are presented and discussed in the following. The injection of pure silicon material into plasma provides a safer and less disruptive alternative compared to the use of silane (SiD$_4$) gas, which introduces more deuterium and is both toxic and explosive. This study addresses the question of under what plasma conditions, real-time injection of materials in small pellet form is suitable to produce Si coatings on main plasma-facing components. This technique is promising for renewing coatings on divertor targets, which are subject to stronger erosion than the main wall. Conventional conditioning techniques that rely on glow discharges may also be less effective in depositing coatings on the targets in closed divertor geometries \cite{bortolon_real-time_2019}. The impact on the main plasma will be discussed, and estimates of the Si erosion yields will be provided. In addition, post-mortem material analysis will be used to investigate the surface structure and composition of the materials.
\begin{figure}[ht]
\begin{center}
\includegraphics[width=82mm]{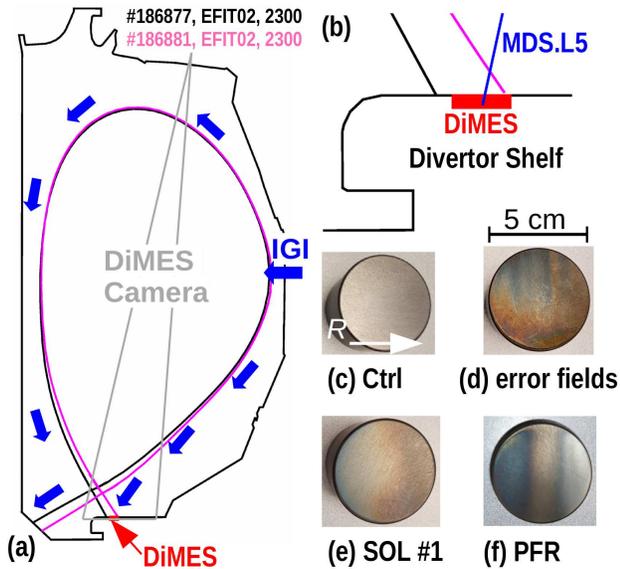}
\caption{\label{fig:figure1}(a) DIII-D poloidal cross-section with magnetic equilibrium separatrices of shots \#186877 (black) and \#186881 (magenta). (b) Zoomed-in view on lower DIII-D divertor shelf, showing outer strike point locations for samples either exposed to the scrape-off layer (SOL) or private-flux region (PFR) at a major radius of $R_{DiMES}=147-152$ cm in using the divertor material evaluation system (DiMES) located at a toroidal angle of 150$^{\circ}$. The blue line indicates the Multi-chord Divertor Spectrometer (MDS) line of sight (L5) on DiMES. (c) DiMES graphite control sample before plasma exposure. (d) Post-exposure sample with a striated footprint in coatings caused by magnetic error fields. (e) A regular Si-rich layer on SOL sample \#1 with error-field correction. (f) Si-rich coating on PFR sample showing the transition from net erosion to net deposition zone. The material samples were exchanged between shots.}
\end{center}
\end{figure}
%%.......................................................................
In section \ref{sec:Setup}, the experimental setup will be explained, followed by an analysis of the suitability of low and high confinement plasma scenarios \ref{sec:Plasmascenario}. Silicon erosion fluxes are considered based on spectroscopic analysis in section \ref{sec:Siliconspectroscopy}. Post-mortem analysis will reveal the surface structure and chemical composition in sections \ref{sec:post-mortem1} and \ref{sec:post-mortem2}, respectively. Lower bound estimates of the deposition and extrapolation for a reactor application are provided in section \ref{sec:post-mortem3}. A summary and conclusions will be provided in section \ref{sec:Conclusion}.  

%-------------------------------------------------------------------------------
\section{The experimental setup for real-time small silicon pellet injection\label{sec:Setup}}
%-------------------------------------------------------------------------------
The impurity granule injector (IGI) is used for high-speed, horizontal injection of macroscopic impurity granules (small pellets) of diameter 0.3-2 mm \cite{vorenkamp_2017}. It is located on the outboard midplane (285-R0 port) of the DIII-D tokamak at a toroidal angle of 285 degrees (figure \ref{fig:figure1}(a)). The IGI consists of a linear, quasi-periodic, granule feeder, which feeds a double-blade rotating impeller. Granules can be injected at 1-150 Hz rates and speeds of 50-200 m/s. However, the injection frequency does not exhibit a regular periodicity due to the decoupled actions of the feeder and the impeller. Therefore, referring to injection rates in the following means an averaged injection frequency typically over 1-2 seconds. The first applications of this small pellet injector were dedicated to ELM-pacing and impurity transport studies on EAST, NSTX-U, and DIII-D \cite{mansfield_2013, bortolon_2016, lunsford_2017, hollmann_2022}. The IGI can inject a variety of materials (carbon, boron carbide, lithium, boron, and other materials) and represents an upgraded version of the original lithium granule injector (LGI)\cite{nagy_2015, vorenkamp_2017}.  

For this study, spherical Si granules (Alfa Aesar, 99.999\% pure) have been injected at 1-15 Hz or mass flow rates of ~1-18 mg/s. The pellet speed is determined by the rotation frequency $f_{imp}$ of the IGI impeller that directs the pellets dropped from the IGI feeder horizontally into the plasma. The pellet velocity can be determined with the impeller blade length $l_p=0.058$ m, $f_{imp}=155$ Hz, and assuming elastic collisions to be $v_p=2\cdot2\pi f_{imp} l_p = 113 $ m/s.

A $\sim$1 mm diameter Si pellet has a mass of $\sim$1.22 mg and deposits $\sim$$2.6\times10^{19}$ Si atoms in the plasma. 
The locations of the magnetic separatrices of two representative magnetic H-mode equilibria are shown in figure \ref{fig:figure1}(a) and (b).

Removable graphite witness samples were exposed to plasmas in the DIII-D open lower divertor using the divertor material evaluation system (DiMES)\cite{rudakov_2017}. The DiMES samples were inserted at $R_{DiMES}=147-152$ cm at 150$^{\circ}$ (figure \ref{fig:figure1}) and exchanged between shots after two plasma exposures.
Langmuir probes embedded in the lower divertor were used to characterize the plasma in the vicinity of DiMES. The IGI flowmeter \cite{vorenkamp_2017}, the extreme ultraviolet (EUV) spectroscopy \cite{fonck_1982}, and the Multi-chord Divertor Spectrometer (MDS) \cite{brooks_1992, abrams_2017} were used to confirm silicon injection and determine the silicon fluxes. The spatial impurity emission distribution in the plasma boundary was monitored with tangential cameras \cite{fenstermacher_tangentially_1997} and vertically with the DiMES TV camera \cite{abrams_2017} using Si II filters (figure \ref{fig:figure1}(a)).

The graphite samples were removed from DiMES post plasma exposure and cut into smaller disks of $\approx$1 cm diameter and 3-4 mm thickness for analysis at Sandia National Laboratories and the Princeton Institute for the Science and Technology of Materials (PRISM). Rutherford Backscattering Spectrometry (RBS), Scanning Electron Microscopy (SEM), Confocal Microscopy, X-ray photoelectron spectroscopy (XPS), Raman spectroscopy, and Fourier-transform infrared spectroscopy (FTIR) were used to resolve the composition and morphology of the sample surface deposits.
%%.......................................................................
\begin{figure}[ht]
\begin{center}
\includegraphics[width=82mm]{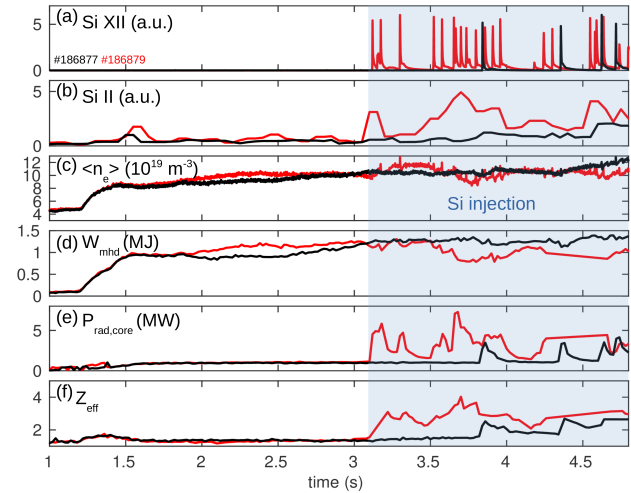}
\caption{\label{fig:figure2} Evolution of H-mode discharges \#186877 and \#186879 with low (black) and high (red) silicon pellet injection rates of ~5 Hz and 15 Hz, respectively: (a) Si XII EUV emissivity, (b) Si II MDS emissivity, (c) line averaged electron density $<n_e>$, (d) stored magnetic energy $W_{mhd}$, (e) total core radiation $P_{rad,core}$, (f) and effective charge $Z_{eff}$. Si injection starts after 3 seconds.}
\end{center}
\end{figure}
%%.......................................................................
%-------------------------------------------------------------------------------
\section{Effects of real-time siliconization on high-confinement plasma performance\label{sec:Plasmascenario}}
%-------------------------------------------------------------------------------
Neutral beam power and density were set to $P_{NBI}= 9.0$ MW and $<n_e>=7.7\times10^{19}$ m$^{-3}$ to allow for sufficiently stable H-mode plasma conditions during Si injection and reduce the risk of disruptions. The toroidal field direction was set to forward $B_t$, which allowed for better error field correction and eliminated striated deposition patterns that would complicate the sample analysis. Si injection rates up to 16 Hz (19.5 mg/s or $4.1\times10^{20}$ Si/s) were achieved in this H-mode scenario. The total radiative power losses increase transiently by about a factor of two during Si injection, while the density shows perturbations on the order of 5-10\%. 
The evolution of the key plasma parameters of two representative H-mode discharges (\#186877 and \#186879) is shown in figure \ref{fig:figure2}(a-f). The Si pellets are injected after 3 seconds at 5 Hz and 15 Hz, respectively. The spikes in the evolution of the EUV Si XII signal show the timing of Si pellet injection. In response, the Si II emission observed by MDS on the DiMES sample increases. Detrimental effects on the core plasma performance manifest for high Si injection rates as a drop in $W_{mhd}$ by up to 35\%, transient increase in the core radiation $P_{rad,core}$ from 1 MW to 4-6 MW, and increase in $Z_{eff}$ from 1.3 to 3-4 during Si injection while such effects remain small for low Si injection rate.                 
%.......................................................................
\begin{figure}
\begin{center}
\includegraphics[width=82mm]{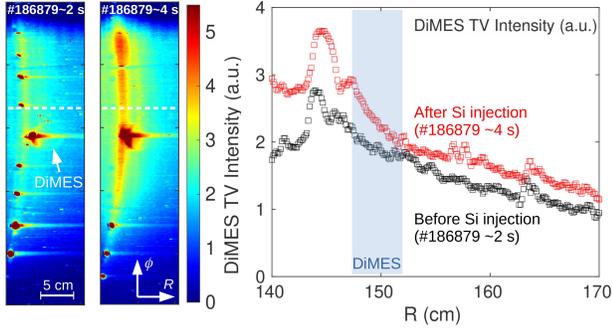}
\caption{\label{fig:figure3} Emission captured from lower outer divertor with the vertical DiMES TV camera view centered around 150$^{\circ}$ using a Si II 636 nm filter for discharge \#186879 (a) before and (b) during 15 Hz silicon pellet injection. Note that the dark red represents areas (mostly continuum emission from leading edges) where the camera detector is saturated. (c) Radial profiles of the emission before and during Si pellet injection along the dashed lines above the DiMES location in (a) and (b).}
\end{center}
\end{figure}
%%.......................................................................
The small silicon pellets were injected during the flat top phase of the discharge. The IGI ablation monitor \cite{vorenkamp_2017} shows emission pulses of pulse width $\Delta t_{abl}=~1$ ms, which represents the time it takes the pellets to ablate fully. Consequently, Si pellets can be estimated to radially penetrate the plasma by a distance $\Delta R_p = v_p\Delta t_{abl}\approx 11.3 $ cm. The distance between LCFS and the wall limiter is 7-7.5 cm for the scenarios considered. Therefore, it can be estimated that the Si pellet penetrates at least 4 cm into the confinement region before it gets fully ablated. The ablated Si eventually diffuses radially outwards into the SOL. Within the SOL, Si is predominantly subject to parallel transport along open field lines and is driven toward the divertor targets by the main ion flow \cite{groth_2007, effenberg_2020}. The combination of radial and parallel transport enables the deposition of injected Si material and the growth of Si-rich films on the plasma-facing components. Figure \ref{fig:figure3} shows a vertical view of the lower divertor and the DiMES graphite sample inserted provided by the DiMES TV camera using a Si II 636 nm filter. The comparison is for discharge \#186879 before and during the 15 Hz Si pellet injection phase. The Si II brightness is substantially increased near the outer strike line, and the DiMES sample exposed to the SOL plasma during silicon injections. The dark red spots represent continuum emission from leading edges at the tile gaps that leak through the filter and tend to cause detector saturation. During a silicon pellet injection, the high-intensity peaks of the Si II line emission resulted in the saturation of the DiMES camera images. The following spectroscopic analysis, therefore, focuses on the gross erosion fluxes between Si pellet injections.
Samples exposed to siliconized plasmas with improved error field correction showed more uniform and regular deposits during post-mortem analysis. The surface of a sample that was exposed to the plasma in the private flux showed a transition between net erosion and net deposition region (figure \ref{fig:figure1}(f)). More details on the surface morphology, chemical composition, and profiles will be discussed in the following sections.
%.......................................................................
\begin{figure}
\begin{center}
\includegraphics[width=70mm]{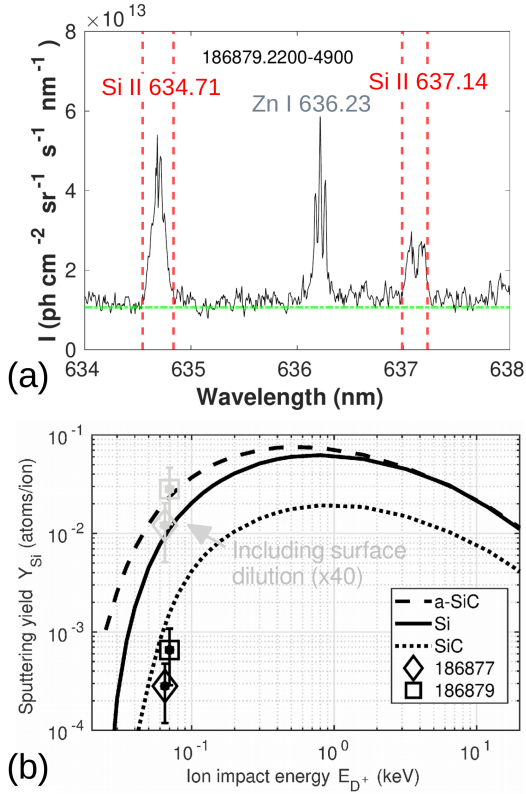}
\caption{\label{fig:figure4} (a) High-resolution MDS emission spectrum used for S/XB analysis to determine Si erosion fluxes and yields in the divertor. (b) Experimentally derived Si erosion yields (symbols) compared to calculated erosion yields for deuterium sputtering on amorphous SiC (dashed), pure Si (solid), and SiC (dotted). The effective erosion yields, adjusted for the dilution effects of Carbon (C) and Oxygen (O) and based on relative concentrations determined using XPS, are shown in grey.}
\end{center}
\end{figure}
%%.......................................................................
During the first sequence of experiments, L-mode discharges (reversed $B_t$, $<n_e>=3.3\times10^{19}$ m$^{-3}$, $P_{NBI}= 2.2$ MW) were established to achieve conditions comparable to recent real-time boronization experiments \cite{effenberg_2020}. Injection of a single Si pellet in these L-mode discharges caused immediate radiative collapse leading to disruption. At a minimum, neutral beam power and density had to be adjusted to $P_{NBI}= 5.5$ MW and $<n_e>=3.95\times10^{19}$ m$^{-3}$ to achieve H-mode plasma conditions that allowed for Si injection at rates of 5-8 Hz. However, those scenarios showed striated patterns in the DiMES TV emission and secondary peaks in the Langmuir probe data. A first sample recovered from the first real-time siliconization experiment showed a toroidally asymmetric lobe-like structured coating on the sample (figure \ref{fig:figure1}(d)). These irregular deposition patterns have been observed on samples during previous real-time boronization experiments in similar plasma scenarios and were found to be caused by the presence of localized perturbations of the magnetic equilibrium that standard error-field correction could not compensate \cite{bortolon_psi_2021, effenberg_2020}. A follow-up study will analyze the cause of the aforementioned error fields and introduce a better compensation method.

The maximum mass injected per discharge was approximately 34 mg of Si. A cumulative amount of 103 mg Si was injected in the shots of interest considered in the following. In total, about 93 small Si pellets ($\approx$114 mg) were injected in all shots, including those neglected for the analysis (a comparative analysis and discussion of the injected and deposited masses within the context of a mass balance analysis will be provided in section \ref{sec:post-mortem3}). A small increase in gas fueling during the first sequence of experiments by 5-10\% suggests a marginal reduction in recycling and an improvement in wall pumping. Good wall conditions are otherwise prevalent due to a recent glow discharge boronization carried out 2-3 weeks before the siliconization experiments.
%.......................................................................
\begin{figure}[ht]
\begin{center}
\includegraphics[width=82mm]{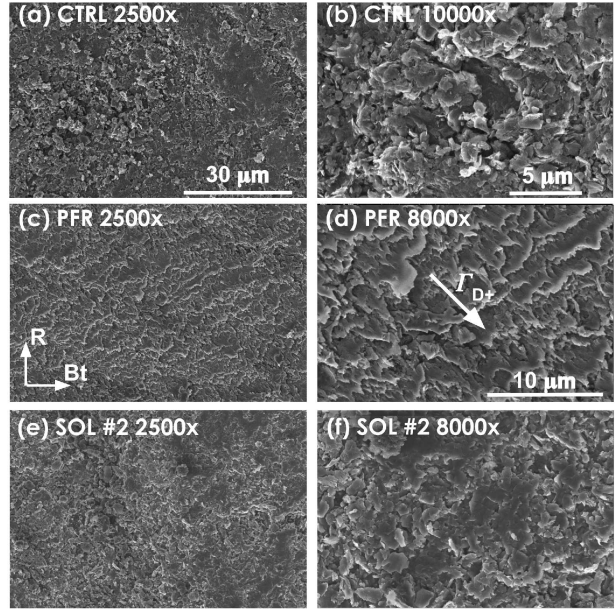}
\caption{\label{fig:figure5} Images from Scanning Electron Microscopy (SEM) for different magnifications: (a) the unexposed graphite control sample at 2500$\times$, and (b) 10000$\times$ magnification, (c) the sample exposed to the private flux region plasma at 2500$\times$ magnification, and (d) 8000$\times$ magnification, (e) the samples exposed to the scrape-off layer plasma at a high Si injection rate at 2500$\times$ magnification, and (f) at 8000$\times$ magnification. The directions of the major radius ($R$), the toroidal magnetic field ($B_t$), and the incoming $D^{+}$ flux ($\Gamma_{D+}$) before entering the sheath are indicated by arrows. The impinging deuterium ion flux $\Gamma_{D+}$ is marked according to the sample orientation with respect to the magnetic field and anticipated ion incident angles based on an equation-of-motion model and measurements
\cite{abe_2022, abe_2022a}}
\end{center}
\end{figure}
%%.......................................................................
\begin{figure*}
\begin{center}
\includegraphics[width=170mm]{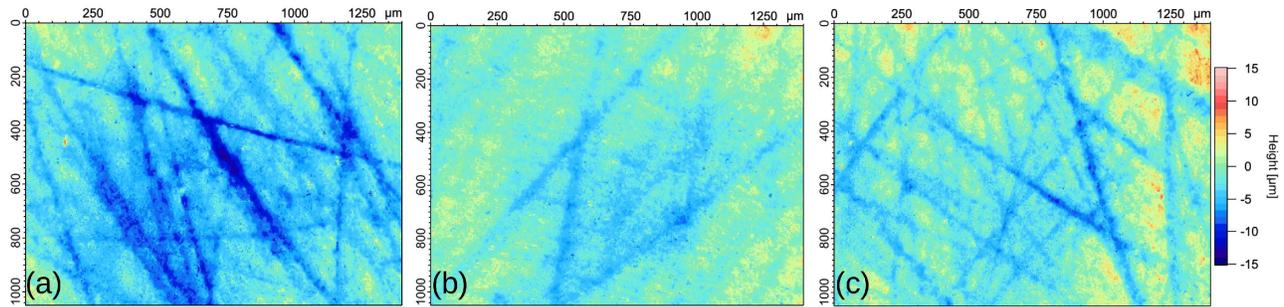}
\caption{\label{fig:figure6} False-color images of the surface topology of DiMES samples from Confocal Microscopy: (a) control sample, (b) sample in private flux region, and (c) sample exposed in SOL (\#2) at high Si injection rate.}
\end{center}
\end{figure*}
%%.......................................................................
%Confocal microscopy image of the control sample topography
%.......................................................................
%-------------------------------------------------------------------------------
\section{Determination of silicon erosion yields using the S/XB method\label{sec:Siliconspectroscopy}}
%-------------------------------------------------------------------------------
The gross erosion rate of Si on the exposed graphite samples was determined between silicon pellet injections using the S/XB method (ionization/photon) \cite{pospieszczyk_2010}. Absolutely calibrated spectroscopic intensities of the Si-II doublet at $634.71/637.14$ nm (figure \ref{fig:figure4}(a)) from MDS were used to calculate removal rates of Si atoms from the sample surface. The Si erosion flux densities were calculated based on this formula\cite{abrams_2021sic}:
\begin{equation*}
    \Gamma_{Si}=\frac{4\pi}{1-r_{Si}}\left(\frac{S}{XB}\right)_{Si-II}I_{Si-II},
\end{equation*}

with the ionization/photon coefficient $\left(\frac{S}{XB}\right)_{Si-II}$ from the ADAS database \cite{summers_2007}, $I_{Si-II}$ the measured spectral Si-II line emission intensity, and the fraction of Si$^{+1}$ ions lost to prompt redeposition $r_{Si}$ before ionization to Si$^{+2}$. The prompt redeposition fraction is calculated based on equation [2] in \cite{naujoks_1996} to be $r_{Si}=0.1-0.2$ for divertor electron densities are $n_{e,div}=1.3-1.4\times 10^{19}$ m$^{-3}$, divertor electron temperatures of $T_{e,div}= 13-14$ eV, and local toroidal magnetic field component of $B_t\approx 2.4$ T. 
Averaged silicon gross erosion fluxes were determined over the time window of 2200 - 4900 ms (slightly broader than the Si injection time window to obtain better statistical averages) for shots $\#$186877 and $\#$186879 with low (5 Hz) and high (15 Hz) Si injection rates using the S/XB method: $\Gamma_{Si, 5 Hz}=5.9\times 10^{14}$ cm$^{-2}$s$^{-1}$ and $\Gamma_{Si, 15 Hz}=1.4\times 10^{15}$ cm$^{-2}$s$^{-1}$. This corresponds to erosion rates of $\sim 0.1$ nm/s.

In order to determine the silicon erosion yields $Y^{*}$, averaged deuterium fluxes of $\Gamma_{D+}\approx2.1\times10^{18}$ cm$^{-2}$s$^{-1}$ were determined based on the ion saturation current $j_{sat}$ from the Langmuir probe closest to the graphite sample, assuming $\Gamma_{D+}=j_{sat}\sin(\theta_{pitch})/e$, with $\theta_{pitch}$ being the pitch angle of the magnetic field lines to the divertor surface. Assuming a sufficiently collisional plasma where the main ion, electron, and impurity ion temperatures are equal, the dependency of sputtering yields on divertor electron temperature, $T_e$, is calculated by considering a 3D Maxwellian ion impact energy distribution shifted by $3ZT_e$ to accommodate the sheath potential drop, with D assigned a value of $Z = 1$ and C assumed to have a value of $Z = 2.5$ based on previous OEDGE/DIVIMP modeling \cite{abrams_2017}. Ion impact energies of 65 eV and 70 eV were extrapolated from the Langmuir probe data. With $\Gamma_{Si}=Y\Gamma_{D+}$ and assuming $E_i=5T_e$, gross erosion yields of $Y=2.8\times10^{-4}$ and $6.6\times10^{-4}$ silicon atoms per deuterium ion are estimated for low and high Si injection rate, respectively. Those values are close to reference values for sputtering yields of D on SiC, as shown for comparison in figure \ref{fig:figure4}(b) (red) but lower than for a pure Si or amorphous silicon carbide (a-SiC). However, the low measured gross erosion yields are likely a result of the surface being diluted with carbon and oxygen (see chemical composition analysis in section \ref{sec:post-mortem2}). This dilution reduces the erosion of Si by introducing additional elements that are preferentially sputtered. Effective erosion yields adjusted for C and O dilution effects based on relative concentrations determined by XPS (section \ref{sec:post-mortem2}) are about a factor $\times40$ higher and show good agreement with the reference data for pure Si or amorphous SiC (grey data points in figure \ref{fig:figure4}). The reference gross erosion yields for Si, SiC, and a-SiC were calculated using the SDTrimSP binary collision approximation code \cite{mutzke_2011} in the static approximation assuming normal incidence \cite{bringuier_2019}. Note that separate treatment of chemical erosion was not possible since no SiD molecular emission was found. The effects of Si sputtering by C ions have also been neglected. The carbon ion flux is 1-2\% of the deuterium ion flux \cite{chrystal_2016, abrams_2018}. The Zn I line at 636.23 nm in figure \ref{fig:figure4}(a) likely results from handling of the sample with latex and nitrile gloves during manufacturing, which may contaminate the sample with traces of zinc \cite{abrams_2021sic}.

Predictive modeling with SDTrimSP was also used prior to the experiments to calculate the expected Si net erosion rate \cite{mutzke_2011, abrams_2021sic}. The calculations used a higher deuterium ion flux of $6.5\times10^{18}$ cm$^{-2}$s$^{-1}$ and suggested Si erosion fluxes to be on the order of $1\times10^{16}$ cm$^{-2}$s$^{-1}$ which corresponds to 1 nm/s. 
The experimental values were lower than the calculated erosion yields by one order of magnitude in the Si erosion fluxes. This is partly due to slightly different plasma parameters in the final H-mode scenarios and assuming erosion from a purely SiC surface in the modeling.
%-------------------------------------------------------------------------------
\section{Morphology analysis of the coated samples\label{sec:post-mortem1}}
%------------------------------------------------------------------------------- 
The first samples exposed to L-mode and H-mode plasmas in a reversed field configuration showed striated deposits due to uncompensated perturbations of the magnetic equilibrium in those scenarios. The effects of error fields were also visible in the Langmuir probe profiles, vertical camera emission profiles, and global deposits on the plasma-facing components. Therefore, those samples were disregarded in the following analysis. Modification of the magnetic configuration and proper error field correction eliminated the perturbations and showed relatively uniform and regularly structured coatings on the samples (compare figure \ref{fig:figure1}(d) and (e)). A coating of varying reflectivity in the radial direction of major radius $R$ was recovered on the sample, which was exposed to the private flux region (figure \ref{fig:figure1}(f)). In this case, the outer strike point was resting on the outer sample edge (compare outer strike line positions in figure \ref{fig:figure1}(a)).
Five small discs have been cut out of each graphite cap for material analysis. Scanning electron microscopy and confocal microscopy were used to analyze the morphology and topographic changes of the surface structure of the sample coatings. In addition, an unexposed sample served as a control sample for comparison during post-mortem analysis. 
SEM results are shown in figure \ref{fig:figure5} for different magnification levels. The comparison is shown for (a) the control graphite sample (CTRL) at 2500$\times$, and (b) 10000$\times$ magnification, (c) with the samples exposed to siliconized plasmas in the private flux region (PFR) at 2500$\times$ and (d) 8000$\times$, and (e) in the scrape-off layer for high injection rates (SOL \#2) at (e) 2500$\times$ and (f) 8000$\times$ magnification. The directions of the major radius ($R$) and toroidal magnetic field ($B_t$) are indicated by arrows. The impinging deuterium ion flux $\Gamma_{D+}$ is marked according to an equation-of-motion model and measurements \cite{abe_2022, abe_2022a}.
\begin{figure}[ht]
\begin{center}
\includegraphics[width=70mm]{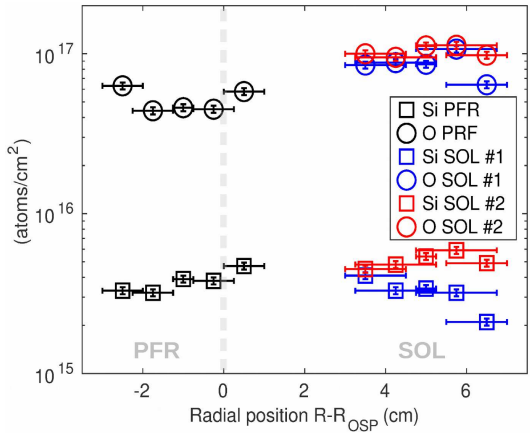}
\caption{\label{fig:figure7} Rutherford Backscattering Spectrometry (RBS) profiles of Si and O surface atomic density on DiMES samples exposed to private flux region and scrape-off layer. The blue and red data represent SOL samples exposed to low and high Si injection rates. The dashed line indicates the outer strike point position.}
\end{center}
\end{figure}
\begin{figure}[ht]
\begin{center}
\includegraphics[width=70mm]{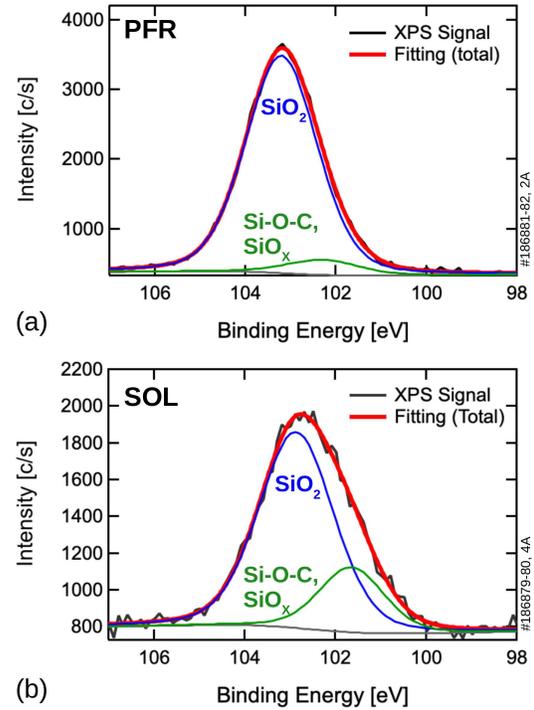}
\caption{\label{fig:figure8} X-ray photoelectron spectroscopy (XPS) spectra for (a) sample 2A (inserted through DiMES during in \#186881-82) in the private flux region (PFR) and (b) sample 4A (inserted through DiMES during in \#186879-80) in the scrape-off layer (SOL). The total fit function (red) can be decomposed into two dominating peaks. The primary peak (blue) is associated with SiO$_2$, the secondary peak (green) represents Si-O-C or SiO$_X$.}
\end{center}
\end{figure}
The surface of the control sample depicted in figure \ref{fig:figure5}(a,b) consists of significant carbon macroscale particles, likely remnants from sample fabrication and polishing. The PFR sample shows a homogeneous surface morphology likely due to erosion in figure \ref{fig:figure5}(c,d). The directionality of the surface features likely reflects the angle of incident ions to the PFR surface.
For the SOL \#2 sample, the reduction in the areal density of macroscale particles (compared to the control sample) indicates the removal of the surface by erosion. The directional surface features are scars from grinding and polishing of the sample prior to plasma exposure. The SOL \#2 sample shows a more heterogeneous surface than the PFR sample. Areas of the surface exhibiting fewer features are likely zones of erosion. The surface features on the SOL \#2 most clearly represent the competing erosion and deposition processes on the sample.

Confocal microscopy was used to determine the surface roughness and height differences. Images of the surface structure of $1400\times1045$ $\mu$m$^2$ size locations on the control sample, on the PFR sample close to the outer strike point, and the SOL sample (\#2) with the highest Si injection rate (15 Hz) are shown in \ref{fig:figure6}(a-c). The dark blue regions ($-15$ to $-10$ $\mu$m) represent deep scratches induced during sample polishing before plasma exposure and remain post-plasma exposure. Root mean square (RMS) roughness values of 2.7 $\mu$m and $2.0-3.5$ $\mu$m were measured for the control sample and the samples under plasma exposure, respectively. The comparison of the different topologies in figure \ref{fig:figure6} suggests a flattening surface post plasma exposure due to erosion in the region closer to the strike line (b). 
The varying topologies presented in the SEM and CM images in figure \ref{fig:figure5} and \ref{fig:figure6} indicate a noticeable morphological change on the surface, specifically in the form of fish scales, along the x-y direction (parallel to the surface). The RMS roughness values demonstrate that the Si deposition and erosion are in the nanometer range and have only minimal impact on $\mu$m structures (table \ref{tab:table1}). While the presence of pore structures or dust particles of larger sizes may augment the RMS roughness, the base surface could concurrently exhibit flattening due to the interplay of erosion and deposition, as suggested in \cite{abrams_2021sic}. It is important to note that the main (micro-)roughness observed is most likely primarily due to the manufacturing and polishing process conducted during sample preparation. The RMS roughness data provide supporting evidence for this claim.

%%.......................................................................
\begin{table*}[t]
  \centering
\begin{tabular}{ |p{1.2cm}|p{1.2cm}||p{1.3cm}|p{1.25cm}|p{1.5cm}|p{1.5cm}|p{1.0cm}|p{1.0cm}|p{1.2cm}|p{1.5cm}|  }
% \hline
% \multicolumn{5}{|c|}{Overview on post-mortem samples} \\
 \hline
 Sample ID& Shot ID 1868XX &Si pellet injection rate (Hz) & RMS roughness ($\mu$m) & Si area density ($10^{15}$ cm$^{-2}$)& O area density ($10^{16}$ cm$^{-2}$)& $Y$ ($10^{-4}$ Si/ion) & Si erosion rate (nm/s) & Si depos. rate (nm/s) & min. Si thickness (nm) \\
 \hline
 PFR       & 81, 82  & $10-12$    & $2.4-2.5$  & $3.2-4.7$   & $4.4-6.3$    & -      & -       &  -         &  $0.6-0.9$ \\
 SOL\#1   & 77, 78  & $4-6$      & $2.0-2.7$  & $2.1-4.1$   & $6.4-10.7$   & 2.8    & 0.1     &  $0.4-0.5$ &  $0.4-0.8$ \\
 SOL\#2   & 79, 80  & $14-16$    & $2.7-3.5$  & $4.5-5.9$   & $9.5-11.3$   & 6.6    & 0.3     &  $0.5-0.7$ &  $0.9-1.2$ \\
 \hline
\end{tabular}
\caption{\label{tab:table1} Overview of samples, shot IDs, Si injection rates, surface roughness, RBS silicon and oxygen areal densities, Si erosion yields, erosion rates, net deposition rates and minimum thickness.}
\end{table*}
%%.......................................................................
%
%-------------------------------------------------------------------------------
\section{Chemical composition analysis of Si-rich deposits\label{sec:post-mortem2}}
%------------------------------------------------------------------------------- 
Post-mortem analysis with Rutherford Backscattering with a 2 MeV $^4$He beam on five spots ($1\times1$ mm) on each sample confirmed the existence of silicon in the in-situ deposits on the graphite samples. Other constituents were carbon, oxygen, and deuterium. The areal Si density was found to be between $2-6\times10^{15}$ Si/cm$^2$. A radial profile is inferred based on the measurement at five locations on the PFR and SOL samples shown in figure \ref{fig:figure7} for silicon and oxygen. The profile data for the SOL have a larger uncertainty concerning the radial position. The Si areal density varies between $2-6\cdot10^{15}$ Si atoms/cm$^{2}$. The oxygen areal density is roughly $10-20$ times magnitude higher than the Si areal density. 
Both O and C can originate from the atmosphere before or after exposure, attributed to the presence of O$_2$, H$_2$O, and CO$_2$. C represents the primary component of the surface (also confirmed by the following XPS analysis), enabling the potential trapping of O through mechanisms like C-O-H.

X-ray photoelectron spectroscopy (XPS) was used to conduct a survey scan for 0-1300 eV (spot size $400$ $\mu$m in diameter) with a factory-set calibration factor using a Thermo Scientific K-Alpha XPS-UPS System, which is based on an Al-Kalpha X-ray source. Bonded Si and pure O peaks were detected at binding energies of 103.05 eV (Si2p) and 533.8 eV (O1s). The bonded Si peaks for the PFR and SOL \#2 samples are shown in figure \ref{fig:figure8}. Pure silicon was not present. Instead, SiO$_2$ and potentially also SiOC and SiO$_x$ were detected on samples exposed to the PFR and the SOL. The relative atomic concentrations were determined by fitting the spectra and found to be 1.2-5\% for bounded Si and 15-24.1\% for pure O. The dominant surface component is C (69-83\%). The prevalence of C composition can explain the observed dilution affecting the erosion discussed in section \ref{sec:Siliconspectroscopy}, further compounded by the presence of O$_2$ with an unknown origin. The highest silicon atomic concentrations were detected near the outer strike point in the PFR. The higher Si concentrations measured on the PFR sample compared to the SOL samples are likely due to the much more significant re-erosion in the latter case. Additionally, electrical drifts in the divertor region affect main ion and impurity ion transport through convective flows \cite{krasheninnikov_1995, moyer_1999}. These $E\times B$ drifts circulate at the boundary between the scrape-off layer and the private flux region, connecting the inner and outer divertor targets. In the forward $B_t$ scenarios considered in the present study, the convective flow generated by the $E\times B$ drift carries silicon impurities from outside the OSP into the PFR. However, a significant Si inflow across the separatrix is also expected from the pedestal region due to the Si pellet penetration depth estimated in section \ref{sec:Plasmascenario}. Recent data from ASDEX-Upgrade experiments with boron powder injection also suggest a substantial material build-up in the private flux region compared to the scrape-off layer caused by $E\times B$ drifts \cite{krieger_2022}.

Fourier-transformed infrared (FTIR) revealed only the presence of carbon-carbon bonds but did not detect Si-C bonding. The divertor target temperatures may not have been sufficient for forming SiC. Usually, SiC formation requires temperatures of 1000 $^o$C and above, but targeted processing efforts brought reported temperatures as low as $\sim500$ $^o$C \cite{hamza_1994, ulrich_1997, dasog_2013}. The experimental conditions were non-ideal for the low-temperature formation of SiC in the present study since the samples were not pre-heated. The surface temperature peaked at about 800 $^{o}$C only at the outer strike point location, according to infrared camera measurements \cite{lasnier_2014} but remained at 200-300$^{o}$ C on the rest of the observed lower divertor target. At lower target temperatures, oxidation processes may dominate. Generally, sticking of incoming oxygen ions, gettering through oxide formation, and air exposure after removal of the samples from the tokamak vacuum chamber may explain the significant oxygen concentration measured according to an analysis of past boronization and siliconization experiments at the tokamak TEXTOR \cite{wienhold_1992, samm_plasma_1995}. In the present study, the oxygen concentrations are generally 4-7 higher on the samples exposed to the siliconized plasmas than the unexposed control sample, suggesting stronger oxygen gettering by the silicon.
%-------------------------------------------------------------------------------
\section{Lower bound estimates for growth rates and thicknesses of divertor Si coatings\label{sec:post-mortem3}}
%------------------------------------------------------------------------------- 
Assuming a pure Si surface layer, a lower bound for the Si coating thicknesses of $0.4-1.2$ nm has been estimated based on the areal density. Taking the discharge evolution and previously estimated erosion rates into account, the net growth rate of the Si-rich coating can be estimated to be at least $0.4-0.5$ nm/s for low Si injection rates and $0.6-0.7$ nm/s for the high Si injection rates. A summary of the surface roughness, RBS areal density, erosion, and deposition data is provided in table \ref{tab:table1}. The erosion rates were found to be $0.1-0.3$ nm/s based on the Si erosion fluxes. Minimum film growth efficiencies of $0.02-0.05$ nm/mg are inferred from the growth and injection rates. Note that the actual coatings may be thicker since bondings with other constituents have been neglected. Given the radial DiMES sample width and the two outer strike point locations, a minimum siliconized area of $A_{Si}\approx 2\pi R \cdot 2\Delta r_{DiMES}=0.94$ m$^2$ can be determined for the outer lower DIII-D divertor. 

Under the assumptions previously discussed, the average mass of deposited silicon (Si) on the outer divertor for each DiMES sample exposed ranges between $1.3$ and $2.3$ mg, which equates to about $4-8$\% of the injected Si mass. For all shots considered (\#186877-82), a total average of 5.25 mg Si was found to have been deposited, representing approximately $5.4$\% of the total Si injected, which was 103 mg. Assuming these conditions, the cumulative thickness of the Si film on $A_{Si}$ might have increased up to $2.4$ nm. This estimate is only based on the radial extent of the DiMES sample surface inserted into the outer lower divertor. The total coated area may be substantially larger since the net deposition zone usually extends into the far SOL beyond the DiMES sample.

Recent modeling with the EMC3-EIRENE and dust injection simulator codes confirmed that the direct deposition of powdered materials was found to be much higher on the inner divertor ($\geq90$\% of injected impurities) compared to the outer divertor \cite{effenberg_2020, effenberg_pfmc_2023}. Those studies also showed that toroidal asymmetries occur in the direct deposition on the divertor when considering toroidally localized impurity sources assuming zero recycling and neglecting re-erosion and re-deposition. However, 1 mm silicon pellets may lead to a more uniform Si flux to the targets due to the pellets' ability to cross the separatrix and subsequent transport in the pedestal/confinement region. Re-erosion and re-deposition may additionally lead to more uniform coatings.

For the applicability of this method in a pilot Fusion Power Plant the significant erosion rates of ~8 mm per year reported in \cite{stangeby_2022} are considered. Extrapolating from the replenishment efficiencies identified in the current study for the outer divertor, it is estimated that 160-400 kg of Si material would need to be injected in the FPP per year to maintain such an 8 mm cladding. With the use of Si powder rather than pellets, higher injection and deposition rates on the order of hundreds of milligrams per second could be attainable in high-performance plasmas. This could lead to more efficient cladding replenishment. In addition, Si deposition on the inner divertor may achieve higher efficiencies and growth rates. The Si deposition dominance in the inner divertor region also suggests that less effort is needed to replenish its PFCs surface coatings compared to the outer divertor. Consequently, through continuous material injection, a cladding thickness of a few millimeters could be effectively maintained in the FPP. While this method may open promising avenues for managing plasma-facing components in future fusion power plants, further research is required to optimize this technique and address issues related to the accumulation of dust and low-Z slag in the main chamber. Controlled experiments and modeling are also needed to broaden the understanding of tritium retention in the exhaust material and the distribution of the coating layers across the reactor.

%-------------------------------------------------------------------------------
\section{Summary and conclusions\label{sec:Conclusion}}
%-------------------------------------------------------------------------------
In-situ growth of silicon-rich films on divertor plasma-facing components has been successfully demonstrated with real-time small Si pellet injection in the DIII-D H-mode scenarios in non-reversed field configurations.

Gross Si erosion yields of $2.8-6.6\times10^{-4}$ and Si erosion fluxes of $0.6-1.4\times10^{15}$ cm$^{-2}$s$^{-1}$ have been determined spectroscopically with the S/XB method. The corresponding Si erosion rates were $0.1-0.3$ nm/s. Surface dilution and the associated preferential sputtering of non-silicon constituents were identified as the primary reasons for the observed low erosion yields.

Si granule injection at 5-19 mg/s created Si-rich layers of at least 0.4-1.2 nm thickness at growth rates of at least 0.4-0.7 nm/s at growth efficiencies of $0.02-0.05$ nm/mg on the outer DIII-D divertor.

The coatings exhibited directional and textured surface morphology consisting of silicon oxides with SiO$_2$ dominating, possibly SiOC, SiO$_X$. The absence of SiC in the coatings was likely due to the low divertor surface temperatures. The sample surface oxygen concentration was found to be approximately a factor of 10 higher than the silicon concentration (in parts possibly due to air exposure of the samples before the analysis).   

For the outer divertor, the Si-rich coatings extend at least over an area of 0.94 m$^2$ covering both the private flux region and scrape-off layer. Deposits may extend over an area of at least 1 m$^2$ on the inner and outer divertor.

Follow-up experiments are suggested with finer Si or SiC powder (e.g., particle size $\leq$ 100 $\mu$m ), which should further reduce the risk of disruptions and allow for more extensive Si injection, leading to increased coating thickness on the main plasma-facing components. Higher divertor target temperatures promote the formation of SiC, which could serve as a replenishable low-mid Z reactor coating material. Additionally, movement of the strike points may allow for the growth and replenishment of coatings on larger surface areas.

A coupling of material transport and surface evolution models is proposed to resolve whether silicon pellet injection helps to equalize deposition levels between targets on the low and high field sides and toroidally \cite{smirnov_2007, schmid_2020, nespoli_2021}. Previous modeling suggested toroidal and poloidal asymmetries with deposition dominating on the high-field side if ablation and ionization occur primarily in the SOL \cite{effenberg_2020, effenberg_pfmc_2023}.

Extrapolations based on the results in this study show that in a pilot Fusion Power Plant, injecting a few hundred kg per year would be necessary to replenish and maintain millimeter-thick low-Z claddings. Less would be needed for the occasional deposition of thinner low-recycling coatings for surface conditioning. Further research is required to optimize the technique, address dust and low-Z slag accumulation, and enhance understanding of tritium retention and coating layer distribution.
%-------------------------------------------------------------------------------
\section*{Acknowledgements}
%-------------------------------------------------------------------------------
This material is based upon work supported by the U.S. Department of Energy, Office of Science, Office of Fusion Energy Sciences, using the DIII-D National Fusion Facility, a DOE Office of Science user facility, under Awards DE-AC02-09CH11466 (PPPL), DE-FC02-04ER54698 (DIII-D), DE-FG02-07ER54917 (UCSD), DE-NA0003525 (SNL), DE-AC52-07NA27344 (LLNL), and DE-AC05-00OR22725 (ORNL). The authors acknowledge the use of Princeton’s Imaging and Analysis Center, which is partially supported through the Princeton Center for Complex Materials (PCCM), a National Science Foundation (NSF)-MRSEC program (DMR-2011750). The DIII-D data shown in this paper can be obtained in digital format by following the links at \url{https://fusion.gat.com/global/D3D_DMP}. 
The author Florian Effenberg gratefully acknowledges discussions with Stefan Bringuier, Anthony Leonard, Nikolas Logan, Adam Mclean, Shawn Zamperini, and \v{Z}ana Popovi\'c.
The United States Government retains a non-exclusive, paid-up, irrevocable, world-wide license to publish or reproduce the published form of this manuscript, or allow others to do so, for United States Government purposes.
%%-------------------------------------------------------------------------------
\section*{Disclaimer}
%-------------------------------------------------------------------------------
This report was prepared as an account of work sponsored by an agency of the United States Government. Neither the United States Government nor any agency thereof, nor any of their employees, makes any warranty, express or implied, or assumes any legal liability or responsibility for the accuracy, completeness, or usefulness of any information, apparatus, product, or process disclosed, or represents that its use would not infringe privately owned rights. Reference herein to any specific commercial product, process, or service by trade name, trademark, manufacturer, or otherwise does not necessarily constitute or imply its endorsement, recommendation, or favoring by the United States Government or any agency thereof. The views and opinions of authors expressed herein do not necessarily state or reflect those of the United States Government or any agency thereof.
%%----------------------------------------------------------------------------
\section*{References}
%%-------------------------------------------------------------------------------

\bibliography{main}
\end{document}